\documentclass[twocolumn,prb,superscriptaddress]{revtex4}
\usepackage{graphicx}
\usepackage{dcolumn}
\usepackage{bm}
\usepackage{amsmath}
\usepackage{times}
\usepackage{epstopdf}
\usepackage{color}
\usepackage[breaklinks=true,colorlinks,citecolor=blue,linkcolor=blue,urlcolor=blue]{hyperref}
\makeatletter

\newcommand{\Rmnum}[1]{\expandafter\@slowromancap\romannumeral #1@}
\makeatother
\graphicspath{{images/}}
\begin{document}

\author{Mahammad Tahir}
\affiliation{Department of Physics, Indian Institute of Technology Kanpur, Kanpur- 208016, India}
\author{Subhakanta Das}
\affiliation{Department of Physics, Indian Institute of Technology Kanpur, Kanpur- 208016, India}
\author{Mukul Gupta}
\affiliation{UGC-DAE Consortium for Scientific Research, University Campus, Khandwa Road, Indore- 452001, India}
 \author{Rohit Medwal} 
\affiliation{Department of Physics, Indian Institute of Technology Kanpur, Kanpur- 208016, India}
\author{Soumik Mukhopadhyay}
\affiliation{Department of Physics, Indian Institute of Technology Kanpur, Kanpur- 208016, India}
\title{Enhancement of spin current to charge current conversion in Ferromagnet/Graphene interface}
\begin{abstract}

The use of graphene in spintronic devices is contingent on its ability to convert a spin current into a charge current. We have systematically investigated the spin pumping induced spin-to-charge current conversion at the Graphene/FM interface and the effect of interface modification through high spin orbit coupling (SOC) material (Pt) as an interlayer (IL) of varying thicknesses by using broadband FMR spectroscopy. The spin mixing conductance is enhanced from $1.66 \times 10^{18}$ m$^{-2}$ to $2.72 \times 10^{18}$ m$^{-2}$ whereas the spin current density is enhanced from 0.135$\pm $0.003 to 0.242$\pm$0.004 MA/m$^{2}$ at the Graphene/FM interface due to the interface modification using high SOC material Pt as an interlayer. The spin current to charge current conversion efficiency turns out to be $\approx 0.003$ nm for the Graphene/FM interface. These findings support the idea that Graphene in combination with high SOC material (Pt) could be a potential candidate for spintronic applications, specifically for spin-torque-based memory applications. 
\end{abstract}
\maketitle
\section{Introduction}
Generation, transportation and detection of pure spin current play a fundamental role in spintronic devices. The primary motivations behind ongoing spintronics research are to achieve, non-volatility~\cite{Fert}, improvement of information processing~\cite{Zutic} and higher integration densities of spin current driven logic and memory devices ~\cite{Urazhdin,Sankey}. When a ferromagnetic (FM) thin film is coupled with a nonmagnetic (NM) thin film, interesting phenomena arise, which include perpendicular magnetic anisotropy (PMA)~\cite{Carcia}, interfacial Dzyaloshinshkii-Moriya interaction (iDMI)~\cite{Ham}, chiral damping~\cite{Jue}, Rashba Edelstein effect (REE) ~\cite{Manchon,Wang} and spin pumping ~\cite{ Tser,Tser2,Tser3}. While most of these phenomena have been attributed to high spin-orbit interactions, the heavy metal (HM) layers being a natural choice for the NM layer. Two-dimensional (2D) materials have recently gained much interest in comparison with the conventional NMs because of their monolayer stability, high spin orbit coupling (SOC), and a lack of inversion symmetry in their crystal structures ~\cite{Feng1}. These unique properties make them excellent candidates for interface magnetism engineering in spintronics.

Pure spin currents from the FM layer to the NM layer across the interface are generated by spin pumping via magnetization precession~\cite{Tser2,Tser3}. Quantitatively, the spin pumping is determined by the interfacial spin mixing conductance ($\mathrm{g}^{\uparrow\downarrow}$) that takes into account the amount of spin pumped, backflow spin currents and the enhanced Gilbert damping $(\alpha)$~\cite{Urban,Tser2}. The enhancement of Gilbert damping is more prominent in NM layers with high spin orbit coupling (SOC) because of the stronger interaction between electron spin and lattice. For spin current to charge current conversion there are two principal mechanisms: inverse Rashba-Edelstein effect (IREE)~\cite{JC,VM} and the inverse spin Hall effect (ISHE)~\cite{Azevedo,Saitoha}. The major difference between these two effects is that IREE is often referred to as an interface effect caused by the significant interfacial SOC and an electric field normal to the surface, induced by the breaking of the inversion symmetry while the spins and electrons are confined in the 2D state~\cite{JC,Shen,Feng}. On the other hand, ISHE arises from the spin-orbit scattering within the bulk of the NM material, with the charge current being transverse to the spin current and spin polarization~\cite{Saitoha,Mosendz}. The basic mechanism can be identified by injecting a pure spin current $\mathrm{J}_\mathrm{s}$. For IREE, the surface charge current density $\overrightarrow{ \mathrm{J_c}}$ is given by:
\begin{equation}
\overrightarrow{ \mathrm{J}_\mathrm{c}} = \mathrm{\alpha}_\mathrm{R}\left(\frac{\mathrm{e}}{\hbar}\right)(\hat{\mathrm{z}}\times\overrightarrow{ \mathrm{J_s}})\label{eq1}
\end{equation}
where $\mathrm{\alpha}_\mathrm{R}$ is the Rashba parameter; $\overrightarrow{ \mathrm{J_s}}$ represents the non-equilibrium spin density caused by spin injection; $\hat{\mathrm{z}}$ represents the unit vector normal to the interface that aligns with the inversion symmetry breaking electric field. ~\cite{Shen,Feng,Cheng}. The induced charge current $\mathrm{J}_\mathrm{c}$ due to ISHE is described by:
\begin{equation}
\overrightarrow{ \mathrm{J}_\mathrm{c}} = \mathrm{\theta}_\mathrm{SH}\left(\frac{\mathrm{2e}}{\hbar}\right)(\overrightarrow{ \mathrm{J}_\mathrm{s}}\times\hat{\sigma})\label{eq2}
\end{equation}
where $\mathrm{J}_\mathrm{c}$ is perpendicular to both the spin current density $\mathrm{J}_\mathrm{s}$ and the spin polarization vector ${\sigma}$. The spin Hall angle $\mathrm{\theta}_\mathrm{SH}$ quantifies the spin current to charge current conversion efficiency~\cite{Hirsch,Sinova}.

Recently, for spin to charge current conversion, single layer graphene (SLG) and semiconducting transition-metal dichalcogenide (TMD) monolayers (MLs) have attracted a great deal of interest. Owing to the low atomic number of carbon, pure graphene has a weak SOC, thus minimising the possibility of applications in spintronics~\cite{Zhu,Feng1,Novoselov,Gao,Axel}. In order to overcome this limitation, multiple approaches have been proposed mainly with the objective of enhancing SOC and local magnetic moments in graphene. These approaches includes (i) functionalization of graphene with tiny doses of adatoms or nanoparticles ~\cite{Gao,Balakrishnan} (ii) fabrication of nanostructures with ferromagnetic gate electrodes~\cite{D} and (iii) ferromagnetic insulator (FMI) or ferromagnetic metal (FM) directly attached to graphene in order to induce magnetic proximity effects~\cite{Mendes,Singh,Ramos}. Moreover, when graphene is synthesized directly on top of semiconducting TMDs, it demonstrates a significant increase in SOC ~\cite{Benítez}. Recently, it was demonstrated that a Py electrode under FMR pumps spin current into SLG ~\cite{Tang}. When a heavy metal layer is deposited on graphene, the SOC strength at the interface may increase, making SLG attached to a FM potential candidate for spintronic applications.

In the present work, we investigate the spin pumping and spin-to-charge current conversion in a heterostructure consisting of Quartz/SLG/$\mathrm{Ni}_{80}\mathrm{Fe}_{20}$ [Py(5 nm)] and the effect of interface modification through high SOC material (Pt) as an interlayer (IL) of varying thickness. We study the effect of interfacial SOC strength using IL thickness variation on the spin pumping efficiency in the SLG by using broadband FMR spectroscopy. The results are interesting for spintronics applications, and add to our understanding of how the interface modification could further enhance the spin current to charge current conversion efficiency.

\begin{figure}
\includegraphics[width=0.99\linewidth]{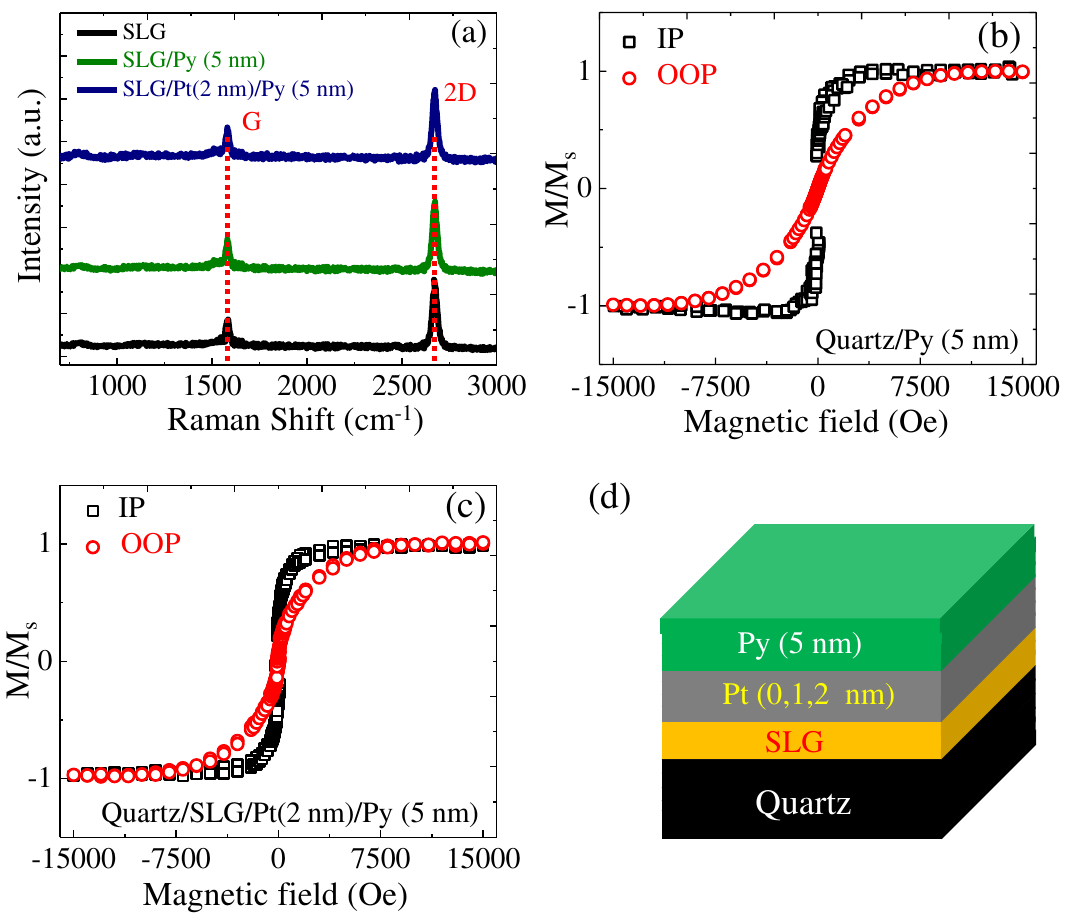}
\caption{(a) Shows the raman spectra of chemical vapor deposition-grown SLG on a Quartz substrate, Quartz/SLG/Py (5 nm), and Quartz/SLG/Pt(2 nm)/Py(5 nm). (b) and (c) In-plane (open black squares) and out-of-plane (open red circles) magnetization hysteresis loops for the stacks involving Quartz/Py(5 nm) and Quartz/SLG/Pt(2 nm)/Py(5 nm), respectively. (d) Illustration of the Quartz/SLG/Pt(0,1,2 nm)/Py(5 nm) stack.}
\label{fig1}
\end{figure}
\begin{figure*}
\includegraphics[width=1.0\linewidth]{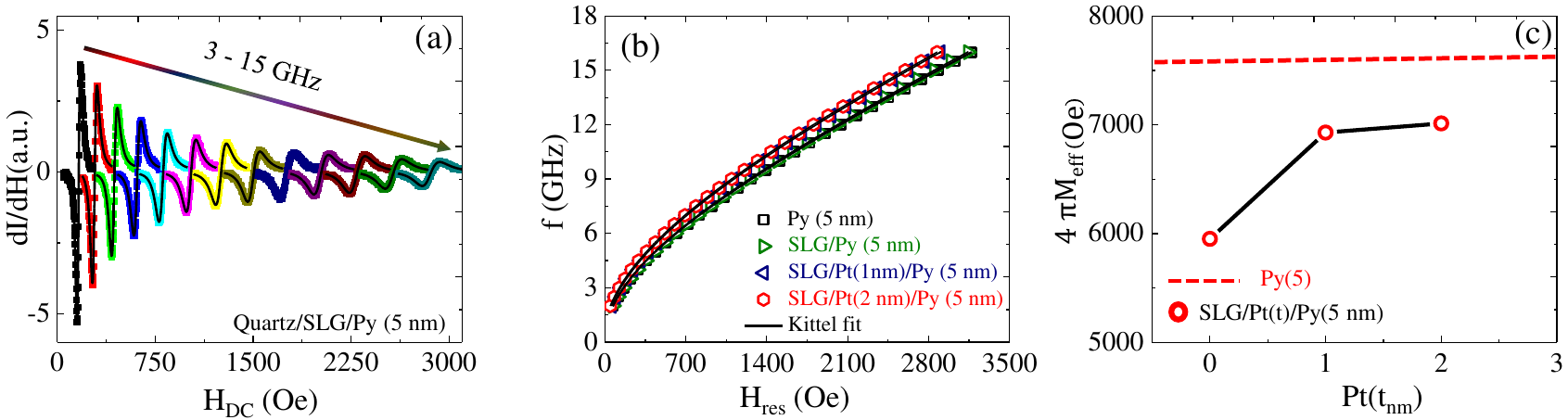}
\caption{(a) FMR spectra recorded on a Quartz/SLG/Py (5 nm) stack. Symbols are experimental data points and black solid line is the fitting using Eq.~\ref{eq3}. (b) Resonance field $\mathrm{H}_{\mathrm{res}}$ vs frequency $\mathrm{f}$ for all the samples. Black lines are the fits using Eq.~\ref{eq4}. (c) Effective magnetization ($4\pi \mathrm{M}_\mathrm{eff}$) for all samples as a function of Pt thickness.}
\label{fig2}
\end{figure*}
\section{Experimental Methods and characterization}

We prepared the thin film heterostructures: Quartz/Py(5 nm), Quartz/SLG/Py(5 nm) and Quartz/SLG/IL/Py(5 nm) with Pt(1,2 nm) as an interlayer (IL). We used commercially available chemical-vapor deposited (CVD) graphene (manufactured by Graphenea) on a Quartz substrate. The thin films of Pt and Py were deposited using DC- magnetron sputtering with a base pressure below 3 × 10$^{-7}$ Torr. The deposition pressure and argon flow rate were 3 × 10$^{-3}$ Torr and 15 standard cubic centimeters per minute (SCCM), respectively. The target was pre-sputtered for two minutes to avoid any contamination. The SLG on Quartz substrate was characterized by Raman spectroscopy (Model: Princeton Instruments Acton Spectra Pro 2500i) equipped with a 532 nm Laser. The saturation magnetization of the thin films was measured by a SQUID-based magnetometer (MPMS 3, Quantum Design). In order to investigate the spin pumping, a lock-in-based broadband ferromagnetic resonance (FMR) set-up (NanOsc) was used. The samples of roughly $4 \times 2$ mm$^2$ were placed on a 200 $\mu$m wide coplaner waveguide (CPW) in a flip-chip manner and the FMR measurements were carried out in the broad frequency range (3–15 GHz) as a function of in-plane external DC magnetic field in a direction perpendicular to the radio frequency field (h$_{\mathrm{rf}}$) at room temperature. For the spin current to charge current conversion measurements, electrical contacts were made using silver paste and gold wire beneath the inverted sample. The spin current to charge current conversion in the in-plane and out of plane geometry were measured by a nanovoltmeter (Keithley 2182A) which is integrated to the FMR setup ~\cite{Tahir,Medwal}.  Out of plane angle-dependent spin current to charge current conversion measurements were performed at the microwave frequencies of 3 and 5 GHz.

The Raman spectra of CVD-grown SLG on a Quartz substrate before and after the deposition of Py (5 nm) and Pt(2 nm)/Py(5 nm) bilayer thin films are shown in Fig.~\ref{fig1}a. The Raman spectra is recorded at a laser wavelength of 532 nm having a spot size of 0.5 $\mathrm{\mu}$m. The low laser power of 125  $\mathrm{\mu}$W is used to avoid any heating effect and thermal damage. The two most commonly recognised SLG signatures are the G peak at $\approx$ 1583.9 $\mathrm{cm}$$^{-1}$ and the 2D (or $\mathrm{G}$$^{'}$) peak at $\approx$ 2677.9 $\mathrm{cm}$$^{-1}$. Here, the 2D peak is identified with second order peak of zone-boundary phonons and the G peak corresponds to a doubly degenerate zone-centre $\mathrm{E}_{\mathrm{2g}}$ mode. We use the spectral weight ratio ($\mathrm{I}$$_{\mathrm{2D}}$/$\mathrm{I}$$_{\mathrm{G}}$) as a quantitative measure of sputtering damage. The $\mathrm{I}$$_{\mathrm{2D}}$/$\mathrm{I}$$_{\mathrm{G}}$ ratio for SLG is 2.07 ± 0.02, which remains unchanged after the deposition of Py (5 nm) and Pt(1,2 nm)/Py (5 nm) bilayer thin films~\cite{Chen}. The typical in-plane (open black squares) and out-of-plane (open red circles) magnetic hysteresis loops for Quartz/Py(5 nm) and Quartz/SLG/Pt(2 nm)/Py(5 nm) stack measured at room temperature are shown in Fig.~\ref{fig1}b and Fig.~\ref{fig1}c, respectively. The saturation magnetization (4$\pi\mathrm{M_{s}}$) value is determined from the magnetic hysteresis measurement for the Quartz/Py (5 nm) sample and is found to be $\approx$ 800 mT while for the Quartz/SLG/Pt (2 nm)/Py (5 nm) bilayer, the value is $\approx$ 750 mT.
The value estimated for effective magnetization from the FMR measurements for the same sample is in good agreement with the stated result.

\section{Magnetization dynamics and spin pumping}
The prepared thin film stacks are subjected to the flip-chip FMR method for magnetization dynamics studies. The measurement schematic with the orientation of the relevant quantities is given in our previous reports ~\cite{Tahir,Medwal}. The FMR measurements are performed in static fields well above the in-plane anisotropy fields so that magnetization $\mathrm{M}$ of the ferromagnetic layer can be considered parallel to the external magnetic field ($\mathrm{H}_\mathrm{ext}$). During magnetization precession, the ferromagnetic system absorbs the applied microwave power, thereby giving rise to the recorded FMR spectrum. The recorded FMR spectra of the Quartz/SLG/Py(5 nm) interface are shown in Fig.~\ref{fig2}a, while the FMR spectra of the Quartz/Py(5 nm) and Quartz/SLG/Pt(1,2 nm)/Py(5 nm) interfaces are shown in Fig.~\ref{fig A1}a and Fig.~\ref{fig A1} (b,c) in the Appendix. The derivative Lorentzian function (Eq.~\ref{eq3}) having symmetric and asymmetric coefficients is used to fit the recorded FMR spectra~\cite{Woltersdorf}.
\begin{widetext}
\begin{equation}
\frac{d \mathrm{I}_{\mathrm{FMR}}}{d \mathrm{H}} = 4\mathrm{A}\frac{\Delta \mathrm{H}( \mathrm{H}- \mathrm{H}_{\mathrm{res}})}{(4( \mathrm{H}- \mathrm{H}_{\mathrm{res}})^2+
(\Delta  \mathrm{H})^2)^2}-\mathrm{S}\frac{(\Delta  \mathrm{H})^2-4(\mathrm{H}-\mathrm{H}_{\mathrm{res}})^2}{(4( \mathrm{H}- \mathrm{H}_{\mathrm{res}})^2+(\Delta  \mathrm{H})^2)^2}\label{eq3}
\end{equation}
\end{widetext}
Here  $\mathrm{H}$, $\Delta \mathrm{H}$ and $\mathrm{H}_{\mathrm{res}}$ are the in plane applied DC magnetic field, FMR linewidth and resonance field of microwave absorption, respectively. The FMR signal's amplitudes $\mathrm{S}$ and $\mathrm{A}$ correspond to symmetric and antisymmetric coefficients, respectively~\cite{Woltersdorf}. We evaluate the $\Delta \mathrm{H}$ and $\mathrm{H}_{\mathrm{res}}$ from the fittings of the FMR spectra. The plots of microwave absorption frequency $\mathrm{f}$  versus  $\mathrm{H}_{\mathrm{res}}$ are shown in Fig.~\ref{fig2}b. The effective magnetization $4\pi \mathrm{M}_{\mathrm{eff}}$ is calculated using Kittel’s equation (Eq.~\ref{eq4})~\cite{Kittel}.

\begin{equation}
\mathrm{f} = \frac{\gamma}{2\pi}\left[\left( \mathrm{H}_{\mathrm{res}}+ \mathrm{H}_{\mathrm{k}}\right) \left( \mathrm{H}_{\mathrm{res}}+ \mathrm{H}_{\mathrm{k}}+ 4 \pi \mathrm{M}_{\mathrm{eff}}\right)\right]^{1/2}\label{eq4}
\end{equation}

Here $\gamma$ = $\frac{g\mu_{B}}{\hbar}$ = $1.88 \times 10^{2}$ GHz/T is the gyromagnetic ratio, $\mu_{B}$ is the Bohr magneton, $g$ is the Lande's spectroscopic splitting factor, $\hbar$ is the reduced Planck's constant, and $\mathrm{H}_{\mathrm{k}}$ is the in-plane anisotropy field of the FM layer. The variation of $4 \pi \mathrm{M}_{\mathrm{eff}}$ as a function of the interlayer Pt thickness is shown in Fig.~\ref{fig2}c. We find that $4 \pi \mathrm{M}_{\mathrm{eff}}$ value is higher for Quartz/Py(5 nm) sample compared to Quartz/SLG/Pt(0,1,2 nm)/Py (5 nm) samples (Fig.~\ref{fig2}c). The value of $4 \pi \mathrm{M}_{\mathrm{eff}}$ is given by  $4 \pi \mathrm{M}_{\mathrm{eff}}$ = $4 \pi \mathrm{M}_{\mathrm{s}}$- $\mathrm{K}_{\mathrm{s}}/\mathrm{M}_{\mathrm{s}}\mathrm{t}_{\mathrm{Py}}$, where $\mathrm{K}_{\mathrm{s}}$ is the surface/interface anisotropy constant, $\mathrm{t}_{\mathrm{Py}}$ is the thickness of the Py layer, and $\mathrm{M}_{\mathrm{s}}$ is the saturation magnetization. Since $\mathrm{K}_{\mathrm{s}}$ is proportional to interfacial spin-orbit coupling, we believe the decrease of $4 \pi \mathrm{M}_{\mathrm{eff}}$ for Quartz/SLG/Pt(0,1,2 nm)/Py(5 nm) is caused by the increase in interfacial spin-orbit coupling due to d-d hybridization, similar to some previous reports for 2-D/FM systems~\cite{Wu,Jamilpanah}. 
\begin{figure*}
\includegraphics[width=1.0\linewidth]{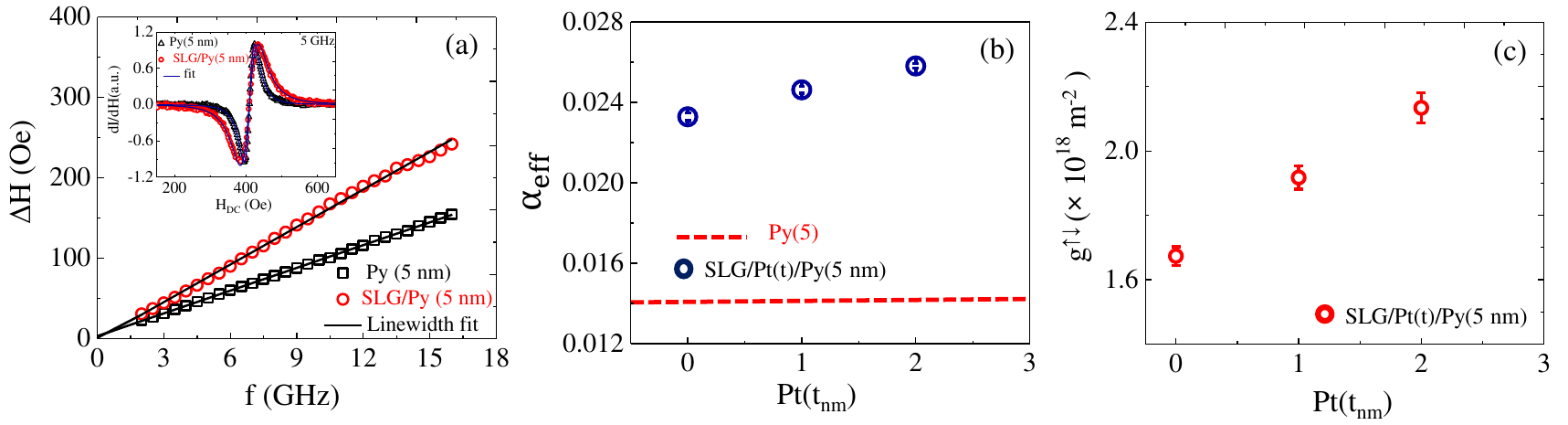}
\caption{(a) Linewidth ($\Delta \mathrm{H}$) versus frequency ($\mathrm{f}$) for Quartz/Py(5 nm) and Quartz/SLG/Py(5 nm) samples. The data points are extrated from FMR fittings, and the solid line are the fitting corresponding  to linewidth equation (Eq.~\ref{eq6}). Inset shows the FMR spectrum of Quartz/Py(5 nm) and Quartz/SLG/Py(5 nm) samples recorded at 5 GHz. Open symbols are the data points while the solid line are the fitting using(Eq.~\ref{eq3}). (b) $\alpha_{\mathrm{eff}}$ for all samples as a function of IL Pt thickness. (c) Interfacial spin mixing conductance ($\mathrm{g}^{\uparrow\downarrow}$) for all samples as a function of IL Pt thickness. }
\label{fig3}
\end{figure*}
The surface/interface imperfections and defects in the FM thin films can enhance the linewidth of the FMR spectra. The total FMR linewidth can be expressed as: 
\begin{eqnarray}
\Delta \mathrm{H}= \Delta\mathrm{H}_
\mathrm{Gilbert}+\Delta\mathrm{H}_\mathrm{inhomogeneous}\label{eq5}
\end{eqnarray}
The Gilbert contributions represent the intrinsic damping. It follows from the Landau Lifshitz-Gilbert equation that the FMR linewidth would depend linearly on the microwave frequency for intrinsic damping. The variation of $\Delta \mathrm{H}$ versus $\mathrm{f}$ are shown in Fig.~\ref{fig3}a. The solid line fit is given by the following equation~\cite{Kittel, Gilbert}:
\begin{eqnarray}
\Delta \mathrm{H}=\frac{4\pi \alpha_{\mathrm{eff}}}{\gamma} \mathrm{f}+\Delta\mathrm{H}_0 \label{eq6}
\end{eqnarray}
Here $\Delta\mathrm{H}_0$ is the frequency-independent contribution known for inhomogeneous broadening, related to the magnetic inhomogeneity in the thin film. The linear behavior of $\Delta \mathrm{H}$ versus $\mathrm{f}$ confirms the intrinsic origin of the damping parameter observed in our Quartz/SLG/Py(5 nm) and Quartz/Py(5 nm) samples, as shown in Fig.~\ref{fig3}a. The FMR spectra of Quartz/Py(5 nm) and Quartz/SLG/Py(5 nm) recorded at 5 GHz are shown in the inset of Fig.~\ref{fig3}a. The value of $\Delta \mathrm{H} = $ 44.58 Oe for Quartz/Py(5 nm) is enhanced to 52.93 Oe at 5 GHz for the Quartz/SLG/Py(5 nm) interface. Spin pumping is responsible for the enhancement in the $\Delta \mathrm{H}$. In the present scenario, the value of $\alpha_{\mathrm{eff}}$ measured from the slope of the linear frequency dependence of linewidth has the following contributions:
\begin{eqnarray}
\alpha_{\mathrm{eff}} = \alpha_{\mathrm{bulk}}+\alpha_{\mathrm{interface}}\label{eq7}
\end{eqnarray}
Here, $\alpha _{\mathrm{bulk}}$ is the damping contribution due to energy transfer to the lattice within the bulk of the ferromagnet~\cite{Hickey}, whereas $\alpha _{\mathrm{interface}}$ is mainly due to the spin pumping which is the result of the spin angular momentum loss due to spin current outflow from the FM layer into the NM layer or due to the spin-flip at the interface due to interfacial spin-orbit coupling~\cite{Conca}.
The value of $\alpha_{\mathrm{eff}}$ is measured as a function of inetrlayer Pt thickness as shown in Fig.~\ref{fig3}b. A large increase in $\alpha _{\mathrm{eff}}$ is observed for the samples Quartz/SLG/Pt(0,1,2 nm)/Py(5 nm) compared to the reference sample Quartz/Py(5 nm) as shown in Fig.~\ref{fig3}b. The enhancement in the Gilbert damping attributed to spin-pumping is estimated by $\Delta\alpha = \alpha_{\mathrm{SLG/Pt(t)/Py}} - \alpha_{\mathrm{Py}}$, which is utilised to calculate the interfacial spin-mixing conductance ($\mathrm{g}^{\uparrow\downarrow}$). The interfacial spin mixing conductance ($\mathrm{g}^{\uparrow\downarrow}$) determines the amount of spin current injected or transmitted by the precessing magnetization vector of the FM layer across the interface and is given by the following expression:~\cite{Tser}
\begin{eqnarray}
\mathrm{g}^{\uparrow\downarrow} =\frac{4\pi \mathrm{M}_{s}\mathrm{t}_{\mathrm{Py}}}{\mathrm{g}\mu_{B}}(\alpha_{\mathrm{SLG/Pt(t)/Py}}-\alpha_{\mathrm{Py}})\label{eqn8}
\end{eqnarray}
where $4 \pi \mathrm{M}_{\mathrm{s}}$ is the saturation magnetization. The extracted values of ($\mathrm{g}^{\uparrow\downarrow}$) at the Quartz/SLG/Py(5 nm) interface is found to be increasing after the interface modification through high SOC material Pt(1,2 nm) as an interlayer. The values of ($\mathrm{g}^{\uparrow\downarrow}$) for Quartz/SLG/Py(5 nm) interface is extracted to be $1.66 \times 10^{18}$ m$^{-2}$ which is further enhanced to $2.72 \times 10^{18}$ m$^{-2}$ for Quartz/SLG/Pt/Py(5 nm) interface as shown in Fig.~\ref{fig3}c.
The enhancement of the Gilbert damping and spin mixing conductance observed in the Quartz/SLG/Pt(0,1,2 nm)/Py(5 nm) stacks could be attributed to: 1) the spin current injected in the SLG by the spin pumping mechanism at the SLG/Pt(0,1,2 nm)/Py(5 nm) interface, which creates a spin accumulation on the SLG. The dissipation of spin current at the SLG/Py and SLG/Pt/Py interfaces through spin-flip scattering acts as an additional channel for spin relaxation, leading to enhanced damping $\alpha_{\mathrm{sp}}$~\cite{Ando1}. The diffusive flow of spins in SLG/Pt(0,1,2 nm)/Py(5 nm) can be described by spin current density $\mathrm{J}_{\mathrm{s}}$, evaluated using the following expression~\cite{Tser,Tser2}:
{\begin{widetext}
\begin{equation}
\mathrm{J}_{\mathrm{s}} \approx \left(\frac{\mathrm{g}^{\uparrow\downarrow}\hbar}{8\pi}\right)\left(\frac{
\mathrm{h}_{\mathrm{rf}}\gamma}{\alpha}\right)^2 \left[\frac{4\pi \mathrm{M}_{\mathrm{s}}\gamma+\sqrt{(4\pi\mathrm{M}_{\mathrm{s}}\gamma)^2+16(\pi \mathrm{f})^2}}{(4\pi\mathrm{M}_{\mathrm{s}}\gamma)^2+16(\pi \mathrm{f})^2}\right]\left(\frac{2e}{\hbar}\right) \label{eq9}
 \end{equation} 
\end{widetext}}
Here $\mathrm{h}_{\mathrm{rf}}$ is the RF magnetic field of $1$ Oe (at 15 dBm rf power) in the strip line of the CPW. The calculated values of $\mathrm{J}_{\mathrm{s}}$ are found to be dependent on interlayer (Pt) thickness and vary within the range of 0.135$\pm $0.003 to 0.242$\pm$0.004 MA/m$^{2}$ at 3 GHz. 
\section{Spin current to charge current conversion}
\begin{figure*}
\includegraphics[width=0.990\linewidth]{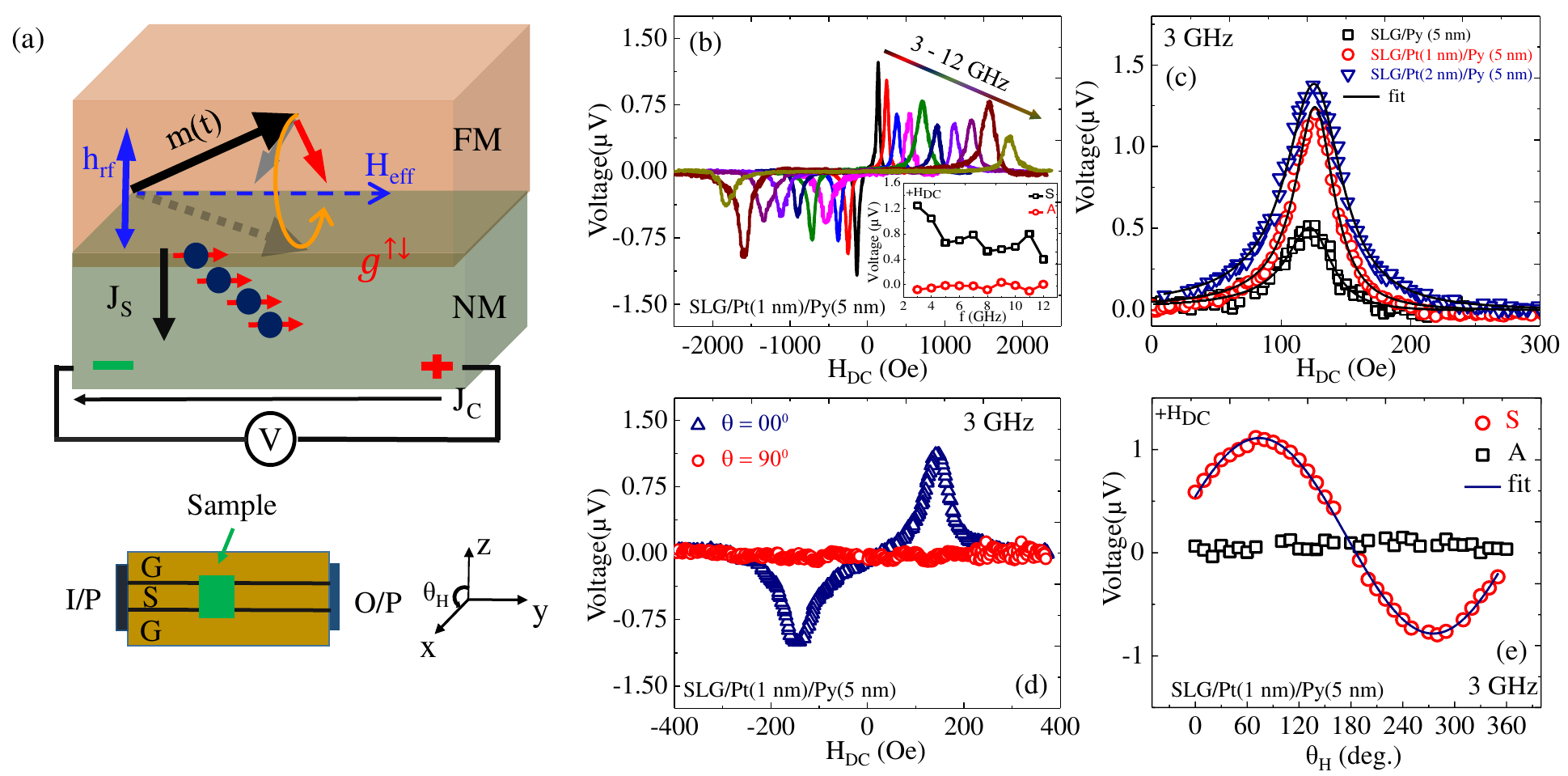}
\caption{(a) The schematic of the FM/NM stack shows how spin pumping facilitates the generation and flow of spin current $\mathrm{J}_{\mathrm{s}}$ across the FM/NM interface. Schematic of CPW showing the thin film stack in contact with the main signal transmission line (S), which is isolated from the adjacent ground line (G). The input and output signal ports of the CPW are represented by I/P and O/P. (b) DC voltages ($\mathrm{V_{dc}}$) as a function of $\mathrm{H_{dc}}$ for SLG/Pt(1 nm)/Py(5 nm) sample for varying frequencies of 3–12 GHz (with step size of 1 GHz) at an excitation power of 15 dBm. Inset shows a symmetric and asymmetric part of voltage with frequency. (c) $\mathrm{V_{dc}}$ with respect to the $\mathrm{H_{dc}}$ for SLG/Pt(0,1,2 nm)/Py(5 nm) samples.(d) DC field scan spin-pumping induced $\mathrm{V_{dc}}$ voltage for f = 3 GHz for sample Quartz/SLG/Pt(1 nm)/Py(5 nm) sample, with the $\mathrm{H_{dc}}$ applied out of the film plane polar angle ($\theta_{\mathrm{H}}$) indicated in the inset.(e) Shows the variation of $\mathrm{V_{sym}}$ and $\mathrm{V_{asym}}$ voltage components with varying out-of-plane polar angle ($\theta_{\mathrm{H}}$) for SLG/Pt(1 nm)/Py(5 nm) interface at an excitation frequency of 3 GHz at rf power of 15 dBm.}
\label{fig4}
\end{figure*}
The study of spin current to charge current conversion is carried out with in-plane microwave excitation. The voltage $\mathrm{V_{dc}}$ is measured with a dc magnetic field ($\mathrm{H_{dc}}$) sweep. The measuring wires are connected to a nanovoltmeter for a direct measurement of the $\mathrm{V_{dc}}$. In Fig.~\ref{fig4}a, the schematic of measurement geometry of the FM/NM stack shows how spin pumping is used for the generation and flow of spin current $\mathrm{J}_{\mathrm{s}}$ across the FM/NM interface. A schematic of CPW shows the thin film stack in contact with the main signal transmission line (S), which is isolated from the adjacent ground line (G). In order to understand the effect of excitation frequency $\mathrm{f}$ on the $\mathrm{V_{dc}}$, we have performed measurements for the Quartz/SLG/Pt(0,1,2 nm)/Py(5 nm) samples in the broad frequency range of 3 to 12 GHz. Fig.~\ref{fig4}b shows the measured $\mathrm{V_{dc}}$ vs $\mathrm{H_{dc}}$ for different $\mathrm{f}$ values for Quartz/SLG/Pt(1 nm)/Py(5 nm) interface. The measured $\mathrm{V_{dc}}$ vs $\mathrm{H_{dc}}$ spectra are symmetric in shape and change sign with inversion of magnetic field direction which indicates that the voltage signal we measure may be primarily due to spin pumping induced spin to charge current conversion in the Pt(0,1,2 nm)/SLG interface. The sign reversal of $\mathrm{{V}_{dc}}$ with reversal of $\mathrm{H_{dc}}$ could be due to the fact that the $\mathrm{V_{dc}}$ $\propto$ $\mathrm{J_s}$$\times$$\sigma$ $\propto$ $\mathrm{J_s}$$\times$$\mathrm{M}$ $\propto$ $\mathrm{J_s}$$\times$ $\mathrm{H}$ $\propto$ $\sin{\theta_\mathrm{H}}$ where the direction of spin polarization $\sigma$ gets reversed upon changing the direction of $\mathrm{H_{dc}}$. 

We envisage the following possibilities regarding the mechanisms responsible for spin-to-charge conversion. 1) The observed $\mathrm{V_{dc}}$ may have contributions of spin pumping induced ISHE voltage ($\mathrm{V_{ISHE}}$) in the Pt(1,2 nm)/SLG interface and spin-rectification effects such as anisotropic magneto resistance (AMR) and anomalous Hall effect (AHE) of the Py layer. The $\mathrm{V_{ISHE}}$ is supposed to generate a symmetric Lorentzian shape, while the AMR/AHE is associated with an asymmetric Lorentzian shape ~\cite{Ando2,Mosendz2}. Therefore, in order to separate the $\mathrm{V_{sym}}$ and $\mathrm{V_{asym}}$ components of voltage we have used the following Lorentzian function to fit the experimental data:
\begin{widetext}
\begin{equation}
 \mathrm{V}(\mathrm{H}) = \mathrm{V}_\mathrm{sym}\frac{(\Delta \mathrm{H})^2}{{( \mathrm{H}- \mathrm{H}_{\mathrm{res}})^2}+(\Delta  \mathrm{H})^2}+\mathrm{V}_\mathrm{asym}\frac{2(\Delta  \mathrm{H})(\mathrm{H}-\mathrm{H}_{\mathrm{res}})}{(\mathrm{H}- \mathrm{H}_{\mathrm{res}})^2+(\Delta\mathrm{H})^2}\label{eq10}
\end{equation}
\end{widetext}
Best fit to the experimental data is shown by a solid black line in Fig.~\ref{fig4}c. Inset of Fig.~\ref{fig4}b shows the $\mathrm{V_{sym}}$ and  $\mathrm{V_{asym}}$ components of voltage measured at 3 to 12 GHz frequencies for SLG/Pt(1 nm)/Py(5 nm) sample. It is clear that $\mathrm{V_{sym}}$ has significant non-zero values for all $\mathrm{f}$, while $\mathrm{V_{asym}}$ is close to zero throughout the frequency range. This confirms that the observed voltage signal $\mathrm{V_{dc}}$ could be due to the ISHE and certainly not AMR/AHE. 

2) Spin to charge current conversion reported for 2D interfaces such as YIG/SLG ~\cite{Mendes}, Ag/Bi ~\cite{JC} and LAO/STO ~\cite{Soumyanarayanan} suggests an alternative mechanism in the presence of 2D SOC driven by broken inversion symmetry. In the present case of SLG/Pt(1,2 nm)/Py(5 nm) interface, we propose that initially the spin current diffuses into the Pt layer with the spin diffusion length ${\lambda}_{\mathrm{N}}$. Subsequently, it is converted into a charge current density shown in Eq.~\ref{eq2} by means of ISHE. This current density leads to a voltage as:
{\begin{widetext}
\begin{equation}
    \mathrm{V}_\mathrm{SP}(\mathrm{H})=\frac{\mathrm{\omega}\, \mathrm{R_N}\, \mathrm{e} \,\theta_\mathrm{SH} \,\lambda_\mathrm{N}\, \mathrm{w} \,\mathrm{P}\, \mathrm{g}_\mathrm{eff}^{\uparrow \downarrow}}{8\mathrm{\pi}} \times \tanh\left({\frac{\mathrm{t_N}}{2\lambda_\mathrm{N}}} \right) \left({\frac{\mathrm{h}}{\Delta\mathrm{H}}} \right)^2 \mathrm{L}(\mathrm{H}-\mathrm{H_R})\, \cos{\phi}\label{eq11}
\end{equation}
\end{widetext}} 
Here $\mathrm{R_N}$, $\mathrm{t_N}$, and $\mathrm{w}$ are, respectively, the resistance, thickness and width of NM layer, and $\mathrm{P}$ is a ellipticity factor and $\omega = 2\pi f$~\cite{Azevedo}. The reported values of spin diffusion length ($\lambda_{\mathrm{N}}$) and spin Hall angle ($\theta_{\mathrm{SH}}$) for Pt are 3.4 nm and 0.056 respectively~\cite{Sanchez}.
\begin{table}[h!]
    \caption{ The values of spin current density ({$\mathrm{J}_\mathrm{s}$}), charge current density ($\mathrm{J}_\mathrm{c}$) and  spin current to charge current conversion efficiency $\mathrm{J}_\mathrm{c}$/{$\mathrm{J}_\mathrm{s}$} calculated for SLG/Pt(0,1,2 nm)/Py(5 nm) samples at 3 GHz.}
    \label{tab:table1}
    \begin{tabular}{cccccc}
{Sample details} &$\mathrm{{J}_s}(\mathrm{MA}/\mathrm{m}^2)$ & {$\mathrm{J}_\mathrm{c}$ ($\mu$A/$\mathrm{m}$)} & $\mathrm{J}_\mathrm{c}$/{$\mathrm{J}_\mathrm{s}$} $(\mathrm{nm}$) \\      
Pt(0 nm) & 0.135 &  0.522 & 0.003\\
Pt(1 nm) & 0.159 &  6.27 & 0.039\\
Pt(2 nm) & 0.242 &  15.12 & 0.062\\
    \end{tabular}
\end{table}
For Py/SLG (Pt=0 nm) interface, we consider the limit $\mathrm{t_N}$/$2{\lambda}_{\mathrm{N}}$ $<1$, so that the expression for spin pumping voltage given by Eq.~\ref{eq11} reduces to:
\begin{equation}
    \mathrm{V}_\mathrm{SP}(\mathrm{H})=\frac{\mathrm{f}\,\mathrm{e} \,\mathrm{R_N}\,  \mathrm{t_N}\,\theta_\mathrm{SH} \, \mathrm{w} \,\mathrm{P}\, \mathrm{g}_\mathrm{eff}^{\uparrow \downarrow}}{8} \times \left({\frac{\mathrm{h}}{\Delta\mathrm{H}}} \right)^2 \label{eq12}
\end{equation}

The spin current is pumped from the FM layer into the Py/SLG (Pt=0 nm) interface and converted by the IREE into a charge current density $\mathrm{j_c} = \left(\frac{\mathrm{2e}}{\hbar}\right)$${\lambda}_{\mathrm{IREE}}\mathrm{J_s}$. The measured voltage $\mathrm{V_{SP}}$ is related to this current density $\mathrm{j_c}$ by $\mathrm{V_{SP}} = \mathrm{R_N}\mathrm{w}\mathrm{j_c}$. Taking into account the values of the effective thickness of SLG $\mathrm{t_N} $ = $0.335 \times 10^{-9}$ m~\cite{Ni}, $\mathrm{R_N} = 0.48\, \mathrm{k}\Omega$, $\mathrm{g}^{\uparrow\downarrow} = 1.66 \times 10^{18} $ m $^{-2}$, $\mathrm{w} = 4$ mm and the amplitude of peak voltage $\mathrm{V} = 0.5 \mu$V at 3 GHz for the Py/SLG (Pt=0 nm) interface, we compute the value of IREE coefficient $\mathrm{\lambda}_\mathrm{IREE}\approx 3\times10^{-3}$ nm. The reported values of $\lambda_\mathrm{IREE}$ in SLG for the Py/SLG interface are  0.003 nm, and for the YIG/SLG interface, the value is  0.002 nm ~\cite{Mendes,Mendes2}.

The spin pumping induced spin to charge current conversion voltage $\mathrm{V_{dc}}$ is enhanced for modified interfaces utilising the Pt IL, as compared to Py/SLG (Pt=0 nm) interface (Fig.~\ref{fig4}c). The observed enhancement of the $\mathrm{V_{dc}}$ could be due to the increase of effective spin-orbit coupling as a result of the insertion of high SOC Pt interlayer.

To measure the voltage dependence on the out-of-plane (OOP) polar angle $\theta_{\mathrm{H}}$ we rotate the sample with respect to the magnetic field. Fig.~\ref{fig4}d shows the $\mathrm{V_{dc}}$ spectra observed at $\mathrm{f} = $ 3 GHz, with $\mathrm{H_{dc}}$ at $\theta_{\mathrm{H}}$ = $0^{\circ}$ and $90^{\circ}$ for SLG/Pt(1 nm)/Py(5 nm) interface.
It can be observed that the sign of $\mathrm{V}_{\mathrm{dc}}$ at $\mathrm{\theta_{H}}$ = $0^{\circ}$ gets reversed when $\mathrm{H}_{\mathrm{dc}}$ is applied in the reverse direction and falls below the noise level when $\mathrm{\theta_{H}}$ = $90^{\circ}$ as shown in Fig.~\ref{fig4}d. 
Fig.~\ref{fig4}e shows the variation of $\mathrm{V_{sym}}$ and $\mathrm{V_{asym}}$ voltage components with varying OOP angle ${\theta_\mathrm{H}}$ for SLG/Pt(1 nm)/Py(5 nm) interface at 3 GHz. As already discussed, the $\mathrm{V_{asym}}$ component of voltage arises due to the AMR and AHE while the $\mathrm{V_{sym}}$ is due to the $\mathrm{V_{ISHE}}$ as well as AMR and AHE~\cite{Lustikova}. We find that $\mathrm{V_{asym}}$ is negligible and remains unchanged with variation in ${\theta_\mathrm{H}}$ , while $\mathrm{V_{sym}}$ is significantly larger and varies sinusoidally with $\mathrm{\theta_{H}}$ as shown in Fig.~\ref{fig4}e. This indicates that the voltage signal measured is primarily due to the spin pumping-induced spin current-to-charge current conversion. 

The values of spin current to charge current conversion efficiency for SLG/Pt(0,1,2 nm)/Py(5 nm) interfaces at 3 GHz are shown in Table~\ref{tab:table1}. For the SLG/Py (Pt = 0 nm) interface, the extracted values of spin current to charge current conversion efficiency are $\approx$ 0.003 nm. While for the SLG/Pt(1 nm)/Py interface, the value of spin current to charge current conversion efficiency is found to be $\approx$ 0.039 nm, which is further enhanced to $\approx$ 0.062 nm for the SLG/Pt(2 nm)/Py interface. These results lead us to speculate that at the SLG/Py (Pt = 0 nm) interface, the IREE may be responsible for the spin to charge current conversion, but at the SLG/Pt (1,2 nm)/Py interfaces, both ISHE and IREE contributed to the spin to charge conversion mechanism.

\section{Conclusion}
In summary, we have systematically observed the dc voltage at the SLG/Py (5 nm) interface as a result of spin pumping-induced spin current to charge current conversion by using a coplanar waveguide-based broadband ferromagnetic resonance setup. We have observed the effect of the interlayer on spin current to charge current conversion by inserting a high SOC material, Pt (1,2 nm), at the SLG/Py interface. We demonstrate a strong correlation of measured spin mixing conductance, spin current density, and spin current to charge current conversion efficiency with the varying thickness of the interlayer Pt. We contemplate that for the spin-current to charge-current conversion in our SLG/Py (Pt = 0 nm) interface, the inverse Rashba-Edelstein effect is responsible, whereas in our SLG/Pt (1,2 nm)/Py interface, the inverse spin Hall effect and the inverse Rashba Edelstein effect both contribute to the conversion mechanism. These findings support the idea that graphene in contact with heavy metals may be a potential material for spintronic applications.

\section{Acknowledgements}
M.T. acknowledges the MHRD, Government of India for Senior Research Fellowship. S.M. acknowledges the Department of Science and Technology (DST) Nanomission, Government of India for financial support. R.M. acknowledges Initiation grant, IIT Kanpur (IITK/PHY/2022027), and I-HUB Quantum Technology Foundation (I-HUB/PHY/2023288), IISER Pune  for financial support.

\appendix
\section{ FMR measurements}
\begin{figure}[h!]
\includegraphics[width=1.05\linewidth]{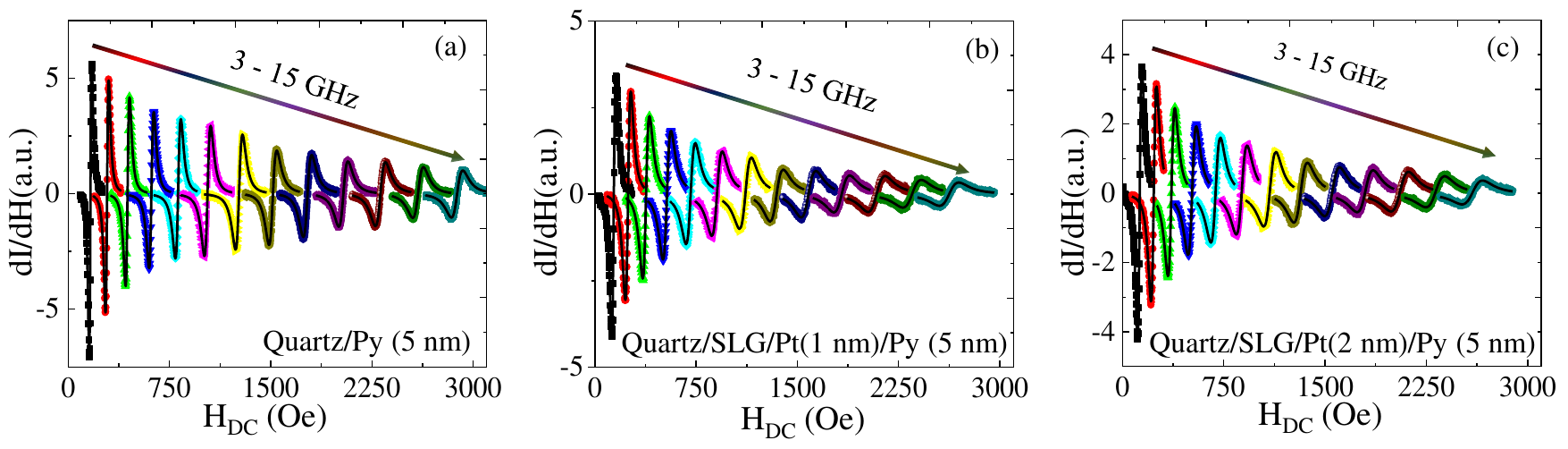}
\caption{Shows the recorded FMR spectra, symbols are experimental data points while black solid line is the fitting using Eq.~\ref{eq3}. (a) Quartz/Py (5 nm) (b) Quartz/SLG/Pt(1 nm)/Py (5 nm) (c)  Quartz/SLG/Pt(2 nm)/Py (5 nm).}
\label{fig A1}
\end{figure}

Fig.~\ref{fig A1} shows the recorded FMR spectra of Quartz/Py(5 nm) stack and Quartz/SLG/Pt(1,2 nm)/Py(5 nm). 
\section{Measurement of spin current to charge current conversion in SLG/Pt(2 nm)/Py(5 nm) interface}
\begin{figure}[h!]
\includegraphics[width=1.05\linewidth]{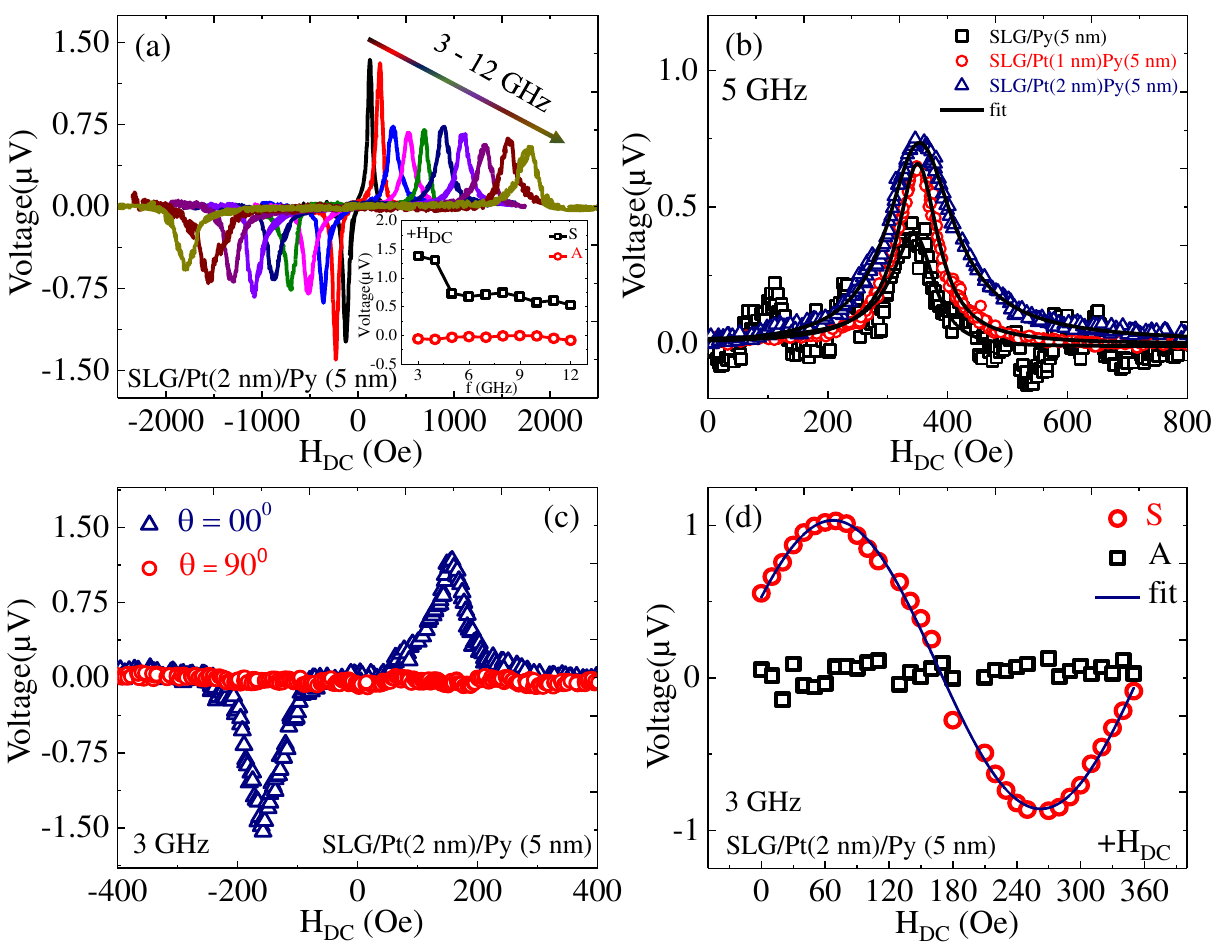}
\caption{(a) DC voltages ($\mathrm{V_{dc}}$) as a function of $\mathrm{H_{dc}}$ for SLG/Pt(2 nm)/Py(5 nm) sample for varying frequencies of 3–12 GHz (with step size of 1 GHz) at an excitation power of 15 dBm. Inset shows a symmetric and asymmetric part of voltage with frequency. (b) $\mathrm{V_{dc}}$ with respect to the $\mathrm{H_{dc}}$ for SLG/Pt(0,1,2 nm)/Py(5 nm) samples at 5 GHz. (c) DC field scan spin-pumping induced $\mathrm{V_{dc}}$ voltage for frequency f = 3 GHz for sample Quartz/SLG/Pt(2 nm)/Py(5 nm), with the  $\mathrm{H_{dc}}$ applied out of the film plane at the polar angle ($\mathrm{\theta}_{\mathrm{H}}$) indicated in the inset (d) Shows the variation of $\mathrm{V_{sym}}$ and $\mathrm{V_{asym}}$ voltage components with varying out-of-plane polar angle ($\theta_{\mathrm{H}}$) for SLG/Pt(1 nm)/Py(5 nm) interface at an excitation frequency of 3 GHz at rf power of 15 dBm.
}
\label{figA2}
\end{figure}
Fig.~\ref{figA2}a shows the measured $\mathrm{V}_{\mathrm{dc}}$ vs $\mathrm{H}_{\mathrm{dc}}$ for different $\mathrm{f}$ values for Quartz/SLG/Pt(2 nm)/Py(5 nm) sample. As we already discussed in the main text that the sign of $\mathrm{V}_{\mathrm{dc}}$ gets reversed when $\mathrm{H}_{\mathrm{dc}}$ is applied in the reverse direction. The measured $\mathrm{V}_{\mathrm{dc}}$ vs $\mathrm{H}_{\mathrm{dc}}$ spectra are symmetric in shape and change sign with inversion of magnetic field direction which indicates that the voltage signal we measure  primarily due to spin pumping induced spin to charge current conversion in the SLG/Pt(2 nm)/Py(5 nm) interface. The spin pumping induced spin to charge current conversion voltage $\mathrm{V_{dc}}$ is enhanced for modified interfaces utilising the Pt IL, as compared to Py/SLG interface. The observed enhancement of the $\mathrm{V_{dc}}$ in Quartz/SLG/IL/Py (5 nm) interface at f = 5 GHz as shown in Fig.~\ref{figA2}b (compared to Py/SLG) could be due to the increase of effective spin-orbit coupling due to the insertion of high SOC Pt interlayer.
 Best fit to the experimental data is shown by a solid black line in Fig.~\ref{figA2}b. Inset of Fig.~\ref{figA2}a shows the $\mathrm{V_{sym}}$ and  $\mathrm{V_{asym}}$ components of voltage measured at 3 to 12 GHz frequencies for SLG/Pt(2 nm)/Py(5 nm) sample. It is clear that $\mathrm{V_{sym}}$ has significant non-zero values for all $\mathrm{f}$, while $\mathrm{V_{asym}}$ is close to zero throughout the frequency range. This confirms that the observed voltage signal $\mathrm{V_{dc}}$ could be due to the ISHE and certainly not AMR/AHE. 
To measure the voltage dependence on the OOP polar angle $\mathrm{\theta_{H}}$ we rotate the sample with respect to the magnetic field. Fig.~\ref{figA2}c shows the $\mathrm{V}_{\mathrm{dc}}$ spectra observed at $\mathrm{f}$ = 3 GHz, with $\mathrm{H}_{\mathrm{dc}}$ at $\mathrm{\theta_{H}}$ = $0^{\circ}$ and $90^{\circ}$ for SLG/Pt(2 nm)/Py(5 nm) interface. It can be observed that the sign of $\mathrm{V}_{\mathrm{dc}}$ at OOP angle $\mathrm{\theta_{H}}$ = $0^{\circ}$ gets reversed when $\mathrm{H}_{\mathrm{dc}}$ is applied in the reverse direction and falls below the noise level when $\mathrm{\theta_{H}}$ = $90^{\circ}$ as shown in Fig.~\ref{figA2}c. Fig.~\ref{figA2}d shows the variation of $\mathrm{V}_{\mathrm{sym}}$ and $\mathrm{V}_{\mathrm{asym}}$ voltage components with varying OOP angle $\mathrm{\theta_{H}}$ for SLG/Pt(2 nm)/Py(5 nm) sample.

\end{document}